\documentclass[conference]{IEEEtran}
\makeatletter
\def\ps@headings{%
\def\@oddhead{\mbox{}\scriptsize\rightmark \hfil \thepage}%
\def\@evenhead{\scriptsize\thepage \hfil \leftmark\mbox{}}%
\def\@oddfoot{}%
\def\@evenfoot{}}
\makeatother
\usepackage{graphicx}
\usepackage{array}
\usepackage{booktabs}
\usepackage{multirow}
\usepackage{url}
\usepackage{listings}
\usepackage{framed}

\ifCLASSINFOpdf
\else
\fi
\begin{document}
\lstset{basicstyle=\footnotesize}

\title{Inferring Internet AS Relationships\\  Based on BGP Routing Policies}

\author{\IEEEauthorblockN{Vasileios Giotsas}
\IEEEauthorblockA{Department of Computer Science\\
University College London\\
Email: v.giotsas@cs.ucl.ac.uk}
\and
\IEEEauthorblockN{Shi Zhou}
\IEEEauthorblockA{Department of Computer Science\\
University College London\\
Email: s.zhou@cs.ucl.ac.uk}
}

\maketitle

\begin{abstract}
The type of business relationships between the Internet autonomous systems (AS) determines the BGP inter-domain routing.
Previous works on inferring  AS relationships relied on the  connectivity information between ASes.  
In this paper we infer  AS relationships by analysing the routing polices of ASes encoded in the BGP attributes {\tt  Communities} and the {\tt Locpref}. 
We accumulate  BGP  data from RouteViews, RIPE RIS and the public Route Servers in August 2010 and February 2011.
Based on the routing policies extracted from data of the two BGP attributes, we obtain  AS  relationships for 39\% links in our data, which include all links among the Tier--1 ASes and most links between Tier--1 and Tier--2 ASes. 
We also reveal a number of special AS relationships,  namely the hybrid relationship, the partial-transit relationship, the indirect peering relationship and the backup links. These special relationships are relevant to a better understanding of the Internet routing.
Our work provides a profound methodological progress for inferring the AS relationships. 

\end{abstract}


\IEEEpeerreviewmaketitle

\section{Introduction}
In the last two decades there has been a great effort in studying the Internet topology at the autonomous systems (AS) level. A number of topology datasets were collected, various topological properties were discovered and a number of network models were proposed \cite{Oliveira:2007, Roughan:2008, Dhamdhere:2011, Donnet:2007, Zhou:2004}. 

The AS topology graph alone, however, is not enough for studying the Internet inter-domain routing. This is because the business relationships between ASes play a crucial role in the decision process of BGP routing  \cite{rfc4271,Tangmunarunkit:2001}. Without the  knowledge on AS relationships, we cannot determine whether a path on the  topology graph is  valid for BGP routing in practice. It is analogue to the situation where we cannot be sure whether there is a direct train service between London and Frankfurt by just looking at the railway map of Europe. 

Internet research and engineering demand data and knowledge on both the AS topology and the AS relationships. For business reasons, ASes do not want to disclose their relationships. 
In the last decade a number of algorithms have been proposed to infer AS relationships based on the AS topology data \cite{gao1, Subramanian1,Dimitropoulos2,Oliveira2}. Since the topology data only contain the connectivity information between ASes, these algorithms had to use various heuristics. The quality of their results have been questioned.

In this paper we propose  to infer  AS relationships using extra sources of information, namely the BGP {\tt Communities} and {\tt LocPrf} (Local Preference)  attributes \cite{Donnet2}. These BGP attributes encode  the routing policies of ASes and therefore closely  reflect the business relationships between ASes.  Our approach allow us to infer AS relationships in a more direct way with increased certainty.

\section{Background}

Internet inter-domain routing is a collaborative effort between ASes, which interconnect and exchange routing information using the BGP protocol. ASes negotiate contractual agreements to define their business relations and impose technical restrictions on traffic exchange. On the Internet, connectivity does not imply traffic reachability, which is fundamentally determined by the business relationships between ASes. 
The AS business relationships are coarsely divided into three categories.
\begin{enumerate}
	\item Transit relationship, including customer-to-provider (c2p) and provider-to-customer (p2c). It is established when an AS (customer) pays a better-connected AS (provider) to transit traffic with the Internet.
	\item Peering relationship, or peer-to-peer (p2p), which allows two ASes to freely exchange traffic between themselves and their customers to avoid the cost of sending traffic through a provider.
	\item Sibling relationship, which allows two ASes (usually under the same administration) to freely exchange traffic without any cost or routing limitations.
\end{enumerate}

BGP routes are usually exported following the so-called \textit{valley-free} rule \cite{gao1}, i.e. a customer route can be exported to any neighbour, but a route from a peer or a provider can only be exported to customers. Hence, a path (of a series of adjacent AS links) is valley-free if it follows such patterns:
(1) $n\times$c2p + $m\times$p2c; or (2) $n\times$c2p + p2p + $m\times$p2c; where n and m $\geq0$.
The sibling links can be inserted freely without changing the valley-free property of a path. 

The valley-free rule describes a typical routing path that is valid for inter-domain routing. Most valid routing paths are valley-free because they comply with the business interest of ASes, i.e. to minimize operation cost and maximize revenue. 
It should be noted that the valley-free rule is not an enforcement rule. It is observed that a small number of routing paths do not follow this rule.

Most ASes try to hide their business relations. In the last decade researchers have introduced a number of algorithms to infer the AS  relationships\,\cite{gao1,Xia:2004, Subramanian1,Erlebach1, Battista2, Dimitropoulos2, Oliveira2, Weinsberg1}. These algorithms have produced conflicting results. BGP simulations using such data have produced poor results \cite{Muhlbauer:2006,Muhlbauer:2007}.
The existing inference algorithms are  limited by the fact that they  relied on the AS connectivity information (obtained from the BGP {\tt ASPATH} attribute or traceroute data) and therefore they had to predict AS relationships by using various heuristics based on a number of assumptions.

In the following we propose to infer the AS relationships based on the AS routing policies encoded in the BGP {\tt Communities} and {\tt LocPrf}  attributes.

\section{Inferring from the {Communities} Attribute}

\subsection{The {Communities} Attribute}

The {\tt Communities} attribute is an entry in BGP update messages. It contains a series of 32-bit numbers, called the {\tt Communities} values. An AS can define many  {\tt Communities}  values with various meanings, and then use them to tag AS links with additional information. For example some values  directly state the AS relationship of the links, and some are instructions for  other ASes to take an action on traffic engineering. Although the {\tt Communities} attribute is optional, it has become intensively used by ASes to facilitate BGP advertisement and to implement flexible routing  policies\,\cite{Donnet2}. 
The {\tt Communities} attribute data can be extracted from  BGP table dumps and update messages and they are also available from public route servers.

Values of the {\tt Communities} attribute are not standardised. Many ASes explain the meaning of their {\tt Communities} values in their Internet Routing Registry (IRR) records \cite{irr} or in the resources of their Network Operation Centers (NOC). A database of NOC websites can be found in the PeeringDB records \cite{PeeringDB}. 

Figure \ref{fig:communities} shows an entry from a BGP table dump. From the {\tt ASPATH} contains three AS links, namely AS4589--AS15412, AS15412--AS18101 and AS18101--AS45528. The  {\tt Communities} values `4589:***' are determined by AS 4589 to describe the  link with AS 15412. For example from  IRR and NOC we  learn that the {\tt Communities} value 4589:612 encodes the meaning `Route received from a LINX peer',  we can identify the relationship between AS4589 and AS15412 is p2p. Similarly, the {\tt Communities} value 15412:705 corresponds to `Route received from customer', we know the relationship of link AS15412--AS18101 is p2c.

\begin{figure}
{
{\tt
TYPE: TABLE\_DUMP\_V2/IPV4\_UNICAST\\
PREFIX: 1.22.73.0/24\\
FROM: 206.223.115.10 AS4589\\
ORIGIN: IGP\\
ASPATH: 4589 15412 18101 45528\\
NEXT\_HOP: 206.223.115.10\\
COMMUNITY: 4589:2 4589:410 4589:612 4589:14413 15412:604 15412:614 15412:621 15412:705 15412:1431 18101:1344 18101:50120 18101:50420\\
}
}
\caption{Example of an entry from the BGP table dump data. }
\label{fig:communities}
\end{figure}

\begin{center}
\begin{table}[t]
\centering	
               \caption{AS Relationships Inferred From Routing Policies}
		\begin{tabular}{ lrrr}
		\toprule
			
           & \textbf{Aug 2010} & \textbf{Feb 2011} \\
			\hline
	    Number of paths  &18,570,393  &  24,549,355 \\		
			Number of AS links & 111,511 &  116,719\\
			Number of ASes & 33,559 &  38,603\\
			\hline
         Number of inferred links  & 43,155 & 43,821 \\
                  {~~~} Number of ASes & 16,877 & 16,918  \\
         \bottomrule 
		  
		Transit relationship & 25,892 & 26,075 \\
                 Peering relationship  & 17,996 & 18,603 \\
                 Sibling  relationship & 176 & 177 \\
		  \hline
                  Hybrid relationship & 909 & 1,034 \\
                 Indirect peering & 708 & 811\\
                  Partial-transit  relationship & 1,526 & 1,828\\
                  Backup links  & 1,087 & 1,205 \\
 \bottomrule 
                 Inferred from {\tt Communities}  & 36,340 & 38,130 \\
                 Inferred from {\tt LocPrf}  & 12,441 & 12,602 \\
		  \bottomrule 		
		  \end{tabular}
	\label{tab:gtRels}
\end{table}
\end{center} 

\subsection{The   {Communities} Attribute Data}

The RouteViews  \cite{routeviews} and the RIPE RIS \cite{ripe3} projects deploy hundreds of BGP monitors around the globe to collect BGP data.  %
We accumulated daily dumps of BGP tables and update messages from all  monitors deployed by RouteViews and RIPE RIS from 1 -- 31 August 2010 and from 1 -- 28 February 2011, respectively. 

We extract AS adjacency information from the {\tt ASPATH} attribute. 
We filter out (1) the reserved and private AS numbers (i.e.\,23456 and 56320--65535) that should not appear in normal BGP advertisements and (2) path cycles that result from misconfiguration. 
%
%We then obtain  18,570,393 unique AS paths in the August 2010 data, which contain 33,559 AS numbers and 111,511 AS links.
%
Table~1 shows the numbers of unique AS paths, AS links and AS numbers (ASN) obtained from the two monthly datasets.

The BGP data contains many BGP attributes. The {\tt ASPATH} attribute has been used in the passive measurement of the Internet AS topology, where all other BGP attributes data were discarded. Subsequently the existing algorithms  have relied on the AS topology information (the passive measurement or the active measurement based on traceroute data) to predict AS relationships. 
Here we utilise extra information sources provided by the  {\tt Communities} and {\tt LocPrf}  attributes
which encode the AS routing polices and therefore can be used to extract AS relationships.

We extract the {\tt Communities} attribute  from our BGP update messages data. 
We obtain {\tt Communities} attribute values for 3,189  ASes. By mining the IRR and the NOCs, we are able to extract the meaning of {\tt Communities} values for 312 ASes, These are well-connected ASes with large numbers of links, including all Tier-1 ASes and the majority of Tier-2 ASes.  

By analysing the routing polices encoded in the meaning of {\tt Communities} values, we directly identify the AS relationships for tens of thousands of links. The routing policy information also enable us to reveal the following four special types of AS relationships that would not be discovered by existing inference algorithms.

\subsection{The Hybrid Relationship}

The traditional model of AS relationships assumes that two ASes have the same type of relationship for all the underlying physical connections. Hence, it is a 1-to-1 model that assigns one relationship type per AS pair. In reality AS interconnections can be more complex, resulting in a cases where two ASes agree different relationship types for different connections. 

\begin{figure}[t]
	\centering
		\includegraphics[width=0.30\textwidth]{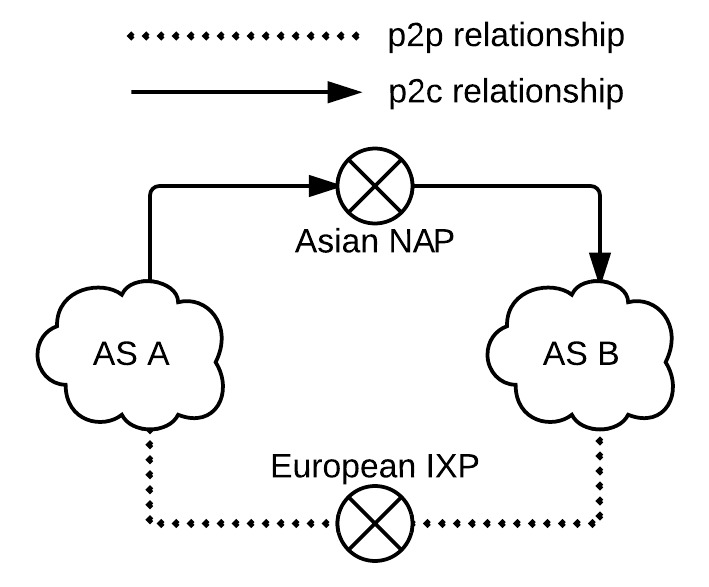}
	\caption{Example of hybrid relationships  between AS A and AS B. The relationship for the AS link through the IXP is p2p, while the relationship for the link through the private NAP is p2c.  }
	\label{fig:hybridRel}
\end{figure}

A hybrid relationship arises when two ASes agree to have both a peering relationship and a transit relationship.
We identify two categories of hybrid links.
 \begin{itemize}
	\item IP version-dependent. Routing policies and paths for IPv4 traffic can differ significantly from those of IPv6. ASes often negotiate separate relationships for prefixes of different IP versions. Therefore two ASes may have a hybrid relationship if they are connected on both  IPv4 and  IPv6 planes.	
	\item Location-dependent. The location of the Points-of-Presence (PoP) can affect AS relationships. Two ASes can have a hybrid relationship when they collocate at more than one  private Network Access Points (NAP) or Internet eXchange Points (IXP).  Figure \ref{fig:hybridRel} shows an example of a location-depended hybrid relationship. 
	\item Some hybrid links are dependent on both IP versions and PoP locations. For instance, two ASes may have an IPv6 transit relationship at a private NAP and an IPv4 peering relationship at an IXP. 
\end{itemize}

A hybrid relationship is identified when a same AS link is tagged with {different} sets  of  {\tt Communities} values in \textit{different} BGP Update messages.
For example, consider the AS link AS3549 -- AS3292. We observe that in a record from a RIPE monitor this link is tagged with the {\tt Communities} 3549:2771 (route received from peer) and 3549:31208 (route received in Denmark), meaning that it is a peering relationship at a connection point in Denmark. Whereas in another record from a RouteViews monitor the same link is tagged with the {\tt Communities} 3549:4354 (route received from customer) and 3549:30840 (route received in the USA), meaning that it is a transit relationship  at a connection point in the USA.

It should be noted that if a link is tagged with different sets of {\tt Communities} values in the {\it same} BGP Update message, we can not conclude  it is a hybrid link. This mainly happens when an AS specifies dual meanings to {\tt Communities} values.
For example, AS1273 uses the  values 1273:1***  to tag customers (where 1*** means all numbers starting with 1) and it uses the values 1273:3*** to tag both providers and route prepending. 
When we observe a link tagged with both   1273:1***  and 1273:3*** in the same BGP record, we can only identify that it  is not a hybrid link but a prepended p2c link  after we learn (from the IRR and NOC data) that prepending {\tt Communities} values are only settable by customers.
Setting dual meanings for a {\tt Communities} value is not a good practice but we observe thousands of such cases in our BGP data. When this happens, we only infer an AS relationship if sufficient extra information is available from other data.

As shown in Table \ref{tab:gtRels}, we discovered that 909 AS links have the transit/peering hybrid relationships in the August 2010 data.
Although a small number, hybrid links are often between well-connected ASes.
We observe that  as high as 13\% of AS links that carry both IPv4 and IPv6 traffic are hybrid links
and more than 10\% of all AS routing paths in our BGP data contain at least one hybrid link.

\subsection{The Indirect Peering Relationship} 

The indirect peering relationship consists of two peering links, which together function as one `virtual' peering link.
It typically occurs when two ASes are peering with the same route server at an IXP such that they gain access to each other's network  as if they have a peering link (without actually having a  physical connection). 
We can detect this indirect peering relationship by the fact that both of the ASes tag the route server as a peering IXP.

Using our {\tt Communities} data collected from BGP update messages, we discover that of the peering links, there are 708 peering links that can form 354 pair of indirect peering relationship. 
Each of the peering link can appear alone in their own routing paths. When two adjacent  peering links form an indirect peering relationship, they do not violate the valley-free principle. From the prospect of Internet  routing, these two peering links can  be replaced by one peering link. 

\begin{figure}
	\centering
		\includegraphics[width=0.30\textwidth]{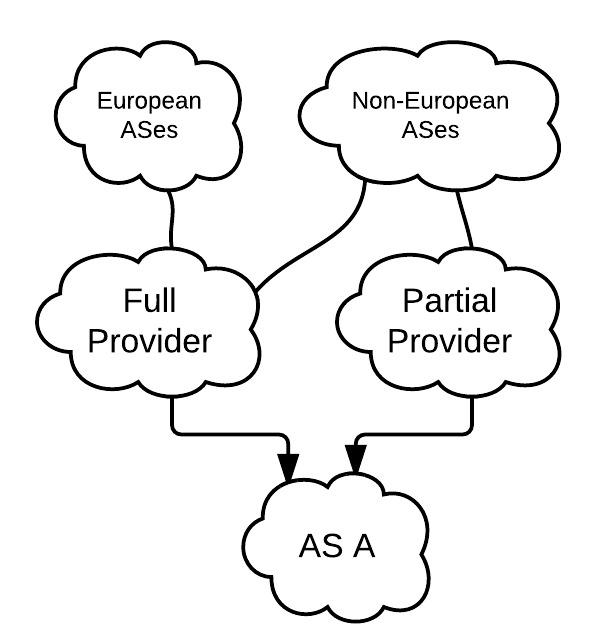}
	\label{fig:partial_transit}
	\caption{Example of the partial-transit relationship. AS A has a partial-transit relationship with the partial provider, which only transit its traffic to non-Europeean ASes. }
	\label{fig:partialRel}
\end{figure}

\subsection{The Partial-Transit Relationship} 

A customer AS  can multihome to more than one providers. 
The partial-transit relationship is a special case of the transit relationship where  providers of a multihomed customer agree to offer transit within  a limited geographical scope. 
A multihomed customer may use  {\tt Communities} values  to instruct a national provider to serve  traffic  destined in the same country and an international provider to serve international traffic  (Figure \ref{fig:partialRel}). 

For example we observe AS3300 (as a provider) provides the customer-settable {\tt Communities}  value 3300:2100 which prevents a customer's route to be announced in Europe.
A partial-transit link is only visible and used locally. Occasionally it can be fully activated (by the customer) if a provider of the customer fails (by setting relevant {\tt Communities} values. 

\subsection{The Backup Links} 

Backup links  are usually invisible and they do not carry any traffic. When there is a disruption in the network, they are activated and  become  visible globally. But they will disappear once the network recovers. The backup links are not a new type of AS relationship. Rather they are {\em transit} links that have the backup function. Backup links are relevant to the Internet routing robustness and reliability. Backup links can be set in the following two ways. In our inference we identify both types of backup links.

When an AS has more than one available routes to a destination, it can set a preferred route and make other routes artificially less favorable.   This is usually achieved by the traffic engineering technique of path prepending. The same technique can be used to create the 
 backup links. The advantage is that such backup links can be automatically activated when the preferred route is disrupted. We identify a prepended backup link if the followings are satisfied: (a) it is a transit link; (b) the customer prepends the {\tt AS\_PATH} attribute such that the link is in an artificially longer path; and (c) we only observe the link for a short lifespan, e.g. less than 5 consecutive days in our monthly data.  
 
Another technique to achieve backup links is the use of the {\tt Communities}  values of {\tt NO-EXPORT} and {\tt NO-ADVERTISE}  that instruct a provider not to advertise the customer routes to anyone. 

\section{Inferring From The {LocPrf} Attribute}
\label{subsec:locpref}

\subsection{The  {LocPrf} Attribute}

An AS with more than one neighbours may receive multiple route advertisements for the same IP prefix. In this case the AS can give each route a preference value, i.e. the {\tt LocPrf} attribute, usually based on the relationship type with the next-hop AS. 
(When the  {\tt LocPrf} attribute cannot determine the best route, other metrics such as the path length are used.)

For a given prefix, the route with the highest {\tt LocPrf} value is used as the preferred route. A usual policy configuration - confirmed by \cite{Wang:2003} - requires that routes received from customers have the highest {\tt LocPrf} value, while routes learned from providers have the lowest value.  
Therefore it is possible to use the {\tt LocPrf} values to reverse-engineer AS relationships. In our inference the {\tt LocPrf} attribute is used as a complementary information source to the  {\tt Communities}  attribute. {\tt Communities} are used for the interpretation of the {\tt LocPrf} values, allowing us to detect exceptions to the above {\tt LocPrf} ordering (e.g. when a peer is given higher preference than a customer).

\subsection{The  {LocPrf} Attribute Data}

{\tt LocPrf} is a local attribute and is not included in the BGP announcements received by RouteViews and RIPE monitors. The {\tt LocPrf} values can be obtained by having a direct interface to a BGP router. Telnet access to such interfaces is provided through public Route Servers that allow remote execution of non-privileged BGP commands.

\begin{figure}%[t]
\centering
	\includegraphics[width=70mm]{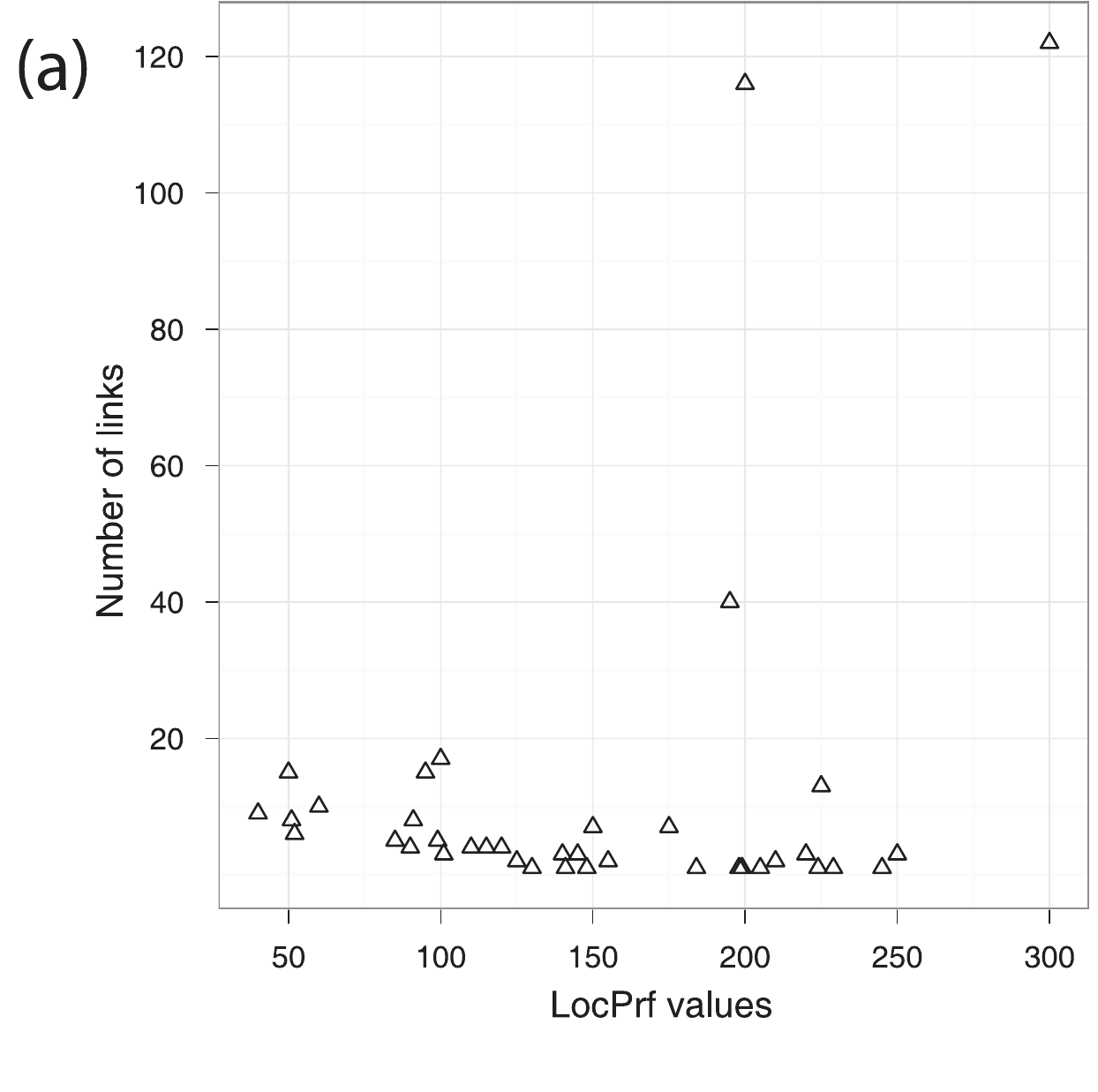}
	
	\includegraphics[width=70mm]{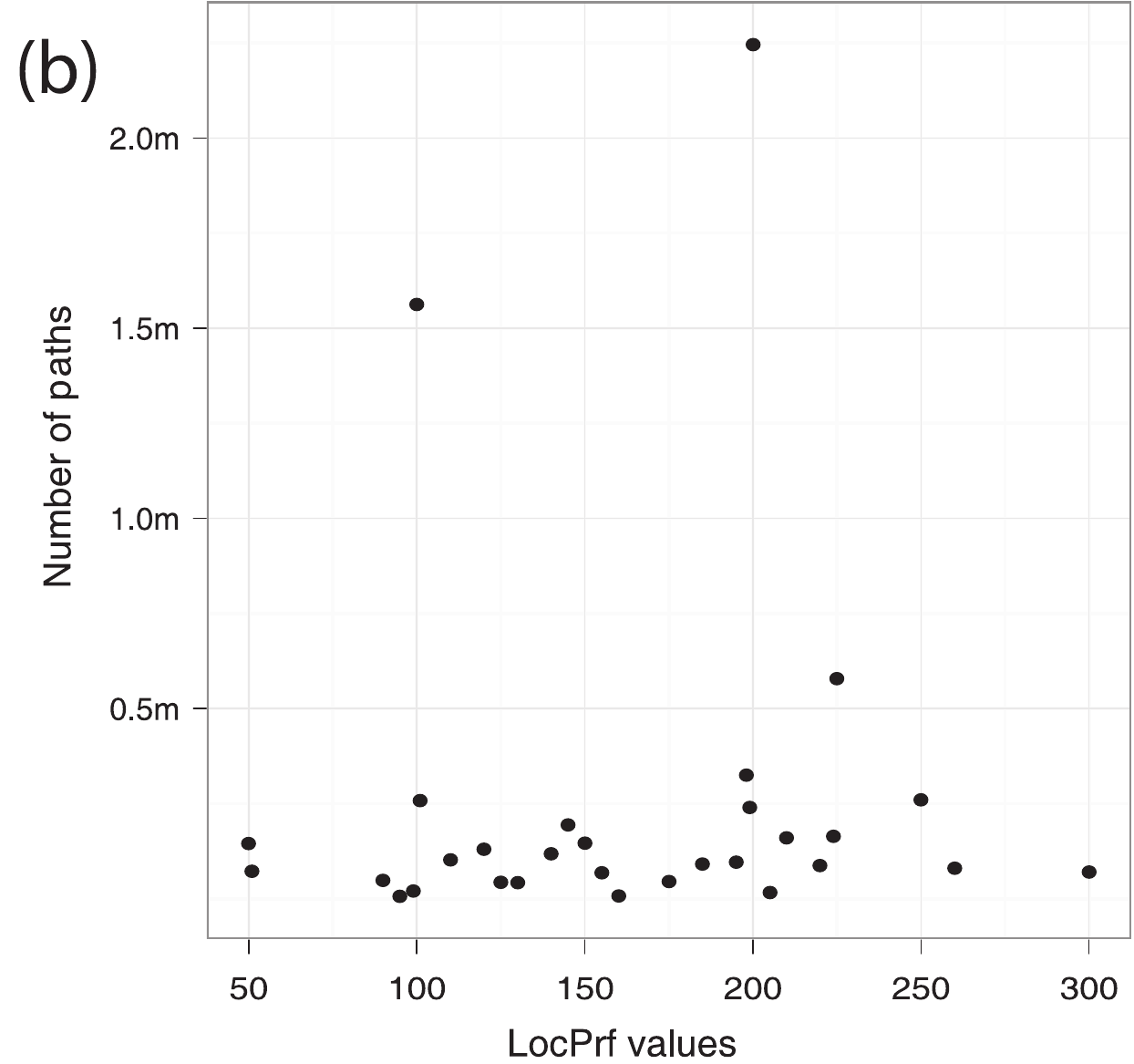}
	\caption{The appearance frequency of the {\tt LocPrf} values of {\tt AS\,4436}  in (a) AS links and (b) AS paths, respectively, in our BGP data. }
	\label{fig:lp4436}
\end{figure}

We  collect weekly table dumps from 28 public Route Servers (that belong to 26 large ISPs) in the same periods of time as above (August 2010 and February 2011 respectively).  
We accumulate 12,441 links which contain 5,839 ASes.

\subsection{Analysing {LocPrf} Attribute Values}

In the simplest case, an AS uses only three {\tt LocPrf} attribute values; the largest value (most preferable) is for the c2p relationship, the smallest value (least preferable) is for the p2c relationship and the middle is for the peering relationship. 

However we observe that most ASes use many  {\tt LocPrf} values. An extreme example is illustrated in Figure\,\ref{fig:lp4436}. 
For example customers can use {\tt Communities}  values to request for upscaling or downscaling their {\tt LocPrf} value for traffic engineering purposes.
For each of such ASes, we try to identify the default {\tt LocPrf} values that are most frequently used:
\begin{enumerate}
	\item For each {\tt LocPrf}   value, we find out the number of links that the AS has assigned the value to. We also search for the number of AS paths in our BGP data that contain these links. We then calculate the distribution of links  and paths, respectively, as a function of  {\tt LocPrf}   values (see Figure\,\ref{fig:lp4436}). 
	\item The {\tt LocPrf} values with the highest frequencies  are chosen as the default values. We may choose more than three default values if their frequencies are significantly larger than  the rest. This happens when two similar {\tt LocPrf} values are  widely used for the same type of relationship with slightly different routing preference. In our work, we have chosen at most 5 default values.	
	\item We  use the meaning of {\tt Communities} attribute values obtained in the above to create a mapping between 
	decide the relationship type of the default  {\tt LocPrf}values. Usually the largest default  value  is for the c2p relationship and the smallest default value is for the p2c relationship. 
\end{enumerate}

In certain cases we can infer the meaning of more  {\tt LocPrf} values based on the default values obtained from the above. For example 
if the majority of prefixes received from a peer AS are tagged with the default peer {\tt LocPrf} value and a few prefixes from the same AS are tagged with a slightly smaller {\tt LocPrf} value, and if this smaller value does not coincide the default transit value, we  conclude that it is also a peer value with a reduced preference. We verify such conclusions against the  {\tt Communities} information and we discard any discrepancy.
Note that the {\tt LocPrf} attribute values can only be used to infer transit and peering relationships.

\section{Our Inference Results}

We combined the inference results obtained from the  {\tt Communities}  and {\tt LocPrf} attributes. As shown in Table~1, we are able to infer  AS relationship for 43,155 links in total, which account for 39\% of the links that are present in our BGP data. These links include all links among the Tier-1 ASes and most links between Tier-1 and Tier-2 ASes.  A hybrid link is counted as both a transit link and a peering link. The partial-transit links and the backup links are included in the total number of transit links. When the relationship of a links is inferred from both BGP attributes, we only accept it if the two reach the same conclusion.

We did not attempt to extract as many AS relationships as possible. 
Rather, our focus is to increase the certainty of the inferred AS relationships. 

\begin{itemize}

\item The  {\tt Communities} and {\tt LocPrf} attributes are configured  by ASes themselves and are  used by them in the BGP routing process. It is expected that ASes should  use them to accurately reflect their business relationships.

\item We collect our BGP  data from the  available sources that have been well studied and widely used. Some of the sources, e.g. the public Route Servers, are playing a crucial role in facilitating the Internet BGP routing.

\item We cross-examine results obtained from different attributes or data sources.  We discard any inconsistency or ambiguity from our results. This sometimes involves large amount of  manual checks.

\item We try to use as few heuristics as possible. When we have to use a heuristic, for example, to identify the default {\tt LocPrf} values, we make sure that the heuristic complies with engineering practice and  supported by previous studies, and we impose  safety checks.

\end{itemize}

We will publish the complete datasets on our webpage at \url{http://web4.cs.ucl.ac.uk/staff/S.Zhou/BAB/}, which will include all raw data and inference results. 

\section{Discussion}

Here we discuss some questions regarding our inference and provide our responses to them. 

\subsection{The meaning of  {Communities}  values}

We extract the meaning of  {\tt Communities}  values mainly from the IRR databases, which may contain inaccurate or stale information \cite{siganos:2004}.
To mitigate this problem, we take out two sanity checks. (1) If an AS has a looking glass or a route server, we cross-check the {\tt Communities} values with {\tt LocPrf} values. (2) If we have the {\tt Communities} values for a link from both sides of the link, we check whether they have the same {\tt Communities} meaning.  Note that there is no incentive for an AS to provide inaccurate {\tt Communities} information, because other ASes use the information to interpret the {\tt Communities} values received from the AS. In our work we found only one AS provided inaccurate {\tt Communities} meanings, which was removed from our study. 

We utilize {\tt Communities} values that not only encode relationship data, but also other policy information such as path prepending or limited route advertisement. This extra policy information provides a valuable resource to understand special relationship types.

\subsection{The meaning of  {LocPrf}  values}

It is rare, but does happen that an AS may use the highest  {\tt LocPrf} value for a peering relationship. We  infer the meaning of a {\tt LocPrf} value not only by comparing the value and its appearance frequency with other values, but more importantly, by using the meaning of {\tt Communities} values tagged on relevant links. We  discard any  {\tt LocPrf} value which is not frequently used or has any uncertainty. Therefore this type of anomalies are excluded from our inference.

\subsection{AS relationship vs BGP routing}
It is rare, but is possible that some ASes implement BGP routing policies that do not exactly reflect their business relationships. It is not a problem, however, if the obtained AS relationship data is used for studying and engineering Internet BGP routing.  

\subsection{Completeness}
We only infer 39\% of links that are visible in our BGP data. The Internet definitely contains  more links than our BGP data. 
Indeed it is  known that many links are missing from the BGP table-based topology measurements \cite{Chang:2006,Oliveira:2008,darkmatter}. 

Nevertheless, our inferred AS relationships cover the majority of  links in the core of  Internet, i.e. links among the Tier-1 ASes and between Tier-1 and Tier-2 ASes.  It is essential to  infer these AS relationships correctly as these links are  important for the global routing. 

The peripheral, small ASes often have only one or two links, which are primarily either a c2p or a s2s links.

Our future work  will try to infer more AS links.

\section{Related Works on AS Relationship Inference}

In the past decade researchers have introduced a number of algorithms to infer AS relationships using  AS topology data.

Based on the valley-free property, Gao \cite{gao1} proposed a relationship inference heuristic that classifies the AS links according to their connectivity degree. Gao's algorithm was refined by a follow-up work \cite{Xia:2004} (PTE) which introduced the use of Partial relationship Information (PI) as a starting point for the inference process.

Subramanian et al \cite{Subramanian1} formulated the ToR as an optimization problem. Two independent studies \cite{Erlebach1, Battista2} proved that the ToR problem is NP-hard and proposed approximate solutions by reducing the ToR to a 2SAT problem which can be solved in linear time. Dimitropoulos et al. \cite{Dimitropoulos2} observed that the ToR formulation can result in multiple solutions without being possible to determine the best. As a response they proposed an enhanced ToR algorithm that incorporates the degree difference as an additional criterion for the maximization of peering links. 

In \cite{Cohen:2007} the Acyclic Type of Relationship (AToR) problem is defined. According to AToR when p2c relationships are assigned a directed edge, the resulting AS graph should be acyclic. In \cite{Hummel:2007} the authors validate the acyclicity of the AS graph and propose a heuristic to solve the maximal AToR problem.

Oliveira et al. \cite{Oliveira2} proposed a more deterministic algorithm exploiting the known fact that the Tier-1 ASes are interconnected with peering relationships. Links that are part of paths that traverse the Tier-1 network are classified as c2p while all the rest are regarded as p2p. Essentially this algorithm is similar to PTE in the sense that the PI is the peering relationships between the Tier-1 ASes, but the extension of the partial information depends solely on the valley-free heuristic. A similar approach is followed in \cite{Weinsberg1} where more general definitions of the Internet core are explored.

These works are common in that they mainly relied on a single data source, i.e. the AS connectivity data. Although sophisticated heuristics have been used, the connectivity data itself is inherently limited in providing useful information for inferring AS relationships. Some  heuristics can even introduce errors. 
Inference results produced by different existing algorithms are often inconsistent and sometimes conflict to each other. 
Two recent works \cite{Muhlbauer:2006,Muhlbauer:2007} showed that BGP simulations based on these data lead to poor results. 
In addition, existing algorithms are not capable of discovering any unconventional AS relationships.

\section{Conclusion}

In comparison to previous algorithms, our approach is  simple and straightforward. 
We  collect the same BGP data in the same way as previous works. The difference is that previous works fundamentally depend on the {\tt ASPATH} attribute, which only contains AS connectivity information. Whereas we look into two other BGP attributes which have been under-utilized by previous works.

ASes use the  BGP attributes {\tt  Communities} and the {\tt Locpref}  
to communicate and implement their routing polices. The attributes data provide valuable information for us to infer AS relationships in a more direct and reliable way.

We do not claim our results are 100\% correct and we intend to make improvements to our method. What is important is that our work provides a profound methodological progress for inferring the AS relationships. 

We only infer 39\% of links visible in our dataset. We did not attempt to infer as many AS relationships as possible.  Rather we make efforts to ensure that the inferred relationships are as accurate and reliable as possible. These links include most of the important links that form the backbone of the Internet. Our future work will aim to infer more AS relationships.

The rich information on the routing polices revealed by the two BGP attributes  allow us to discover a number of special relationship types. The existing  algorithms are incapable of discovering them. These special relationships are relevant to a better understanding of the Internet routing

\section*{ACKNOWLEDGMENT}
S Zhou is supported by The Royal Academy of Engineering and EPSRC under grant number 10216/70.

\bibliographystyle{IEEEtran}
\bibliography{IEEEabrv,infocom}

\end{document}